\documentclass[aps,prd,showkeys,showpacs,amssymb,cite,
amsfonts,epsf,preprintnumbers,nofootinbib,superscriptaddress]{revtex4}

\usepackage[dvips]{graphicx}
\usepackage{bm,latexsym,amsmath,amssymb,amsfonts,color}
\usepackage[mathscr]{eucal}
\newcommand{\be}[1]{\begin{equation} \label{#1}}
\newcommand{\ee}{\end{equation}}
\newcommand{\bea}{\begin{eqnarray}}
\newcommand{\eea}{\end{eqnarray}}
\newcommand{\ba}{\begin{array}}
\newcommand{\ea}{\end{array}}
\newcommand{\bel}{\begin{align}}
\newcommand{\eel}{\end{align}}
\newcommand{\nn}{\nonumber}

\newcommand{\C}{C_{\rm local}}
\newcommand{\A}{\mathfrak{A}}

\begin{document}

\title{Heat capacity of a self-gravitating spherical shell of radiations}
\author{Hyeong-Chan Kim}
\email{hckim@ut.ac.kr}
\affiliation{School of Liberal Arts and Sciences, Korea National University of Transportation, Chungju 380-702, Korea}
\begin{abstract}
We study the heat capacity of a static system of self-gravitating radiations analytically in the context of general relativity. 
To avoid the complexity due to a conical singularity at the center, we excise the central part and replace it with a regular spherically symmetric distribution of matters of which specifications we are not interested in. 
We assume that the mass inside the inner boundary and the locations of the inner and the outer boundaries are given.
Then, we derive a formula relating the variations of physical parameters at the outer boundary with those at the inner boundary. 
Because there is only one free variation at the inner boundary,
the variations at the outer boundary are related, which determines the heat capacity.
To get an analytic form for the heat capacity, we use the thermodynamic identity $\delta S_{\rm rad} = \beta \delta M_{\rm rad}$ additionally, which is derived from the variational relation of the entropy formula with the restriction that the mass inside the inner boundary does not change.
Even if the radius of the inner boundary of the shell goes to zero, in the presence of a central conical singularity, the heat capacity does not go to the form of the regular sphere.
An interesting discovery is that another legitimate temperature can be defined at the inner boundary which is different from the asymptotic one $\beta^{-1}$. 
\end{abstract}
\maketitle

\section{Introduction}
In 1980, Landau and Lifshitz~\cite{Landau1980} pointed out that systems bound by long range forces might exhibit negative heat capacity even though the specific heat of each volume element is positive. 
Since then, many such examples were found, e.g., mainly stars and blackholes.
A self-gravitating isothermal sphere also belongs to this class, which can be regarded as a model of a small dense nucleus of stellar systems~\cite{Lynden-Bell}.   
Based on general relativity, the model was dealt by Sorkin, Wald, and Jiu~\cite{Sorkin:1981wd} in 1981 as a spherically symmetric solution which maximizes entropy.  
Schmidt and Homann~\cite{Schmidt:1999tr} called the geometry a `photon star'. 
The heat capacity and stability of the solution were further analyzed in Refs.~\cite{Pavon1988,Chavanis:2007kn,Chavanis:2001hd,Chavanis:2001ib}. 
Thereafter, the system has drawn attentions repeatedly in relation to the entropy bound~\cite{Schiffer:1989et,Hod:1999as}, blackhole thermodynamics~\cite{Sorkin:1997ja}, maximum entropy principle~\cite{Gao:2016trd,Fang:2013oka,Gao:2011hh}, holograpic principle~\cite{Bousso:2002ju,Lemos:2007ys,Anastopoulos:2013xdk}, and conjecture excluding blackhole firewalls~\cite{Page:2013mqa}. 
A system of self-gravitating radiations in an Anti-de Sitter spacetime was also pursued~\cite{Page:1985em,Vaganov:2007at,Gentle:2011kv}.
Studies on the system of self-gravitating perfect fluids are undergoing~\cite{Pesci:2006sb}. 
An interesting extension of the self-gravitating system was presented in Ref.~\cite{Schmidt:1999tr,Anastopoulos:2011av} where a conical singularity was inserted at the center as an independent mass source from the radiation.
Some of the singular solutions were argued to have an interesting geometry, which is similar to an event horizon in the sense that $1-2m(r)/r$ has a minimum value close to zero.
Analytic approximation was tried to understand the situation that a blackhole is in equilibrium with the radiations~\cite{Zurek}. 
It was also argued that the thermodynamics of a black hole in equilibrium implies the breakdown of Einstein equations on a macroscopic near-horizon shell~\cite{Anastopoulos:2014zqa}.
The geometrical details of solutions having conical singularity were dealt in Ref.~\cite{Kim:2016jfh}.

Let us consider a static spherically symmetric system of self-gravitating radiations confined in a spherical shell bounded by two boundaries located at $r=r_- $ and $r=r_+$ in the context of general relativity.
For a generic time symmetric data, the initial value constraint equations become simply $^{(3)}R = 16\pi \rho$.
As described in Ref.~\cite{Sorkin:1981wd}, this determines the spatial metric to be the form $h_{ij}dx^idx^j= [1-2m(r)/r]^{-1} dr^2 + r^2 d\Omega^2$, where 
\be{mass}
m(r) =M_-+  4\pi \int_{r_-} ^r \rho(r') r'^2 dr'.
\ee
Here $M_-$ represents the mass inside the inner boundary at $r_-$.
At present, we do not assume anything about the nature of $M_-$ except for the spherical symmetry. 
Therefore, it can take negative value.
The mass of the radiations in the shell is 
\be{Mrad}
M_{\rm rad}  =  M_+- M_-; \qquad M_+ \equiv \lim_{r\to \infty} m(r) ,  
\ee
where  $M_+$ denotes the total mass of the solution. 
We neglect the energy density of the confining shell and assume that no matters lie outside of the outer boundary at $r_+$. 
We also assume that the radiation is thermodynamically disconnected with matters inside $r_-$.
The energy density of the radiation at the outer surface of the shell with local temperature $T_+$ is 
\be{rho+}
\rho(r_+) = \sigma T_+^4,
\ee
where $\sigma$ is the Stefan-Boltzmann constant.

Formally, the entropy of the radiation with equation of state, $\rho(r) = 3p(r)$, can be obtained by integrating its entropy density over the volume,~\cite{Sorkin:1981wd}
\be{entropy}
s = \frac{4}{3} \sigma T^3, \qquad
S_{\rm rad}  =  \int_{r_-}^{r_+} L(r) dr; \qquad
L \equiv \frac{4 (4\pi \sigma)^{1/4} }{3}  
	\frac{r^{1/2}[m'(r)]^{3/4} }{\chi(r)} ,
\ee
where $\chi(r) \equiv \sqrt{1- 2m(r)/r}$.
The variation of $S_{\rm rad}$ with respect to a small change of $m(r)$ 
gives 
\bea \label{dSrad}
\delta S_{\rm rad} &=& \int_{r_-}^{r_+} 
\left[ \frac{\partial L}{\partial m} -\frac{d}{dr}\frac{\partial L}{\partial m'} \right] \delta m dr +
	\left[ \frac{\partial L}{\partial m' } \delta m \right]_{r_-}^{r_+} \nn \\
&=& \int_{r_-}^{r_+} 
\frac{\delta S_{\rm rad}}{\delta m}  \delta m dr + \beta_+ \delta M_+ -\beta_- \delta M_-,
\eea
where 
\be{beta}
 \beta_\pm \equiv \left( \frac{\partial L }{\partial m'}\right)_{r\to r_\pm} 
 = \left[\frac{r^{1/2}}{ \chi} \Big(\frac{4\pi \sigma}{m'(r)}\Big)^{1/4} \right]_{r\to r_\pm} .
\ee
Noting the relation of mass with the surface energy density in Eqs.~\eqref{mass} and \eqref{rho+}, the local temperature $T_+$ is related with $\beta \equiv \beta_+$ as
\be{beta+}
\beta^{-1} = \chi_+ T_+ .
\ee
One may introduce the metric component $g_{tt}$ so that the local temperature at $r$ is given by the Doppler-shifted temperature as,
\be{beta}
\sqrt{-g_{tt}(r)}  \,T(r) = \sqrt{-g_{tt}(r)} \Big(\frac{\rho(r)}{\sigma}\Big)^{1/4} = \beta^{-1};
\qquad g_{tt}(r_+) = - g_{rr}(r_+)^{-1},
\ee
where the second condition is introduced so that the metric outside the shell is just the vacuum Schwarzschild solution. 
This result can also be obtained by solving the Einstein's equation directly.
This equation indicates that $\beta^{-1}$ is the global temperature measured at the asymptotic region. 
On the other hand, $\beta_-$ is not directly related with the local temperature $T_-$ by the relation in Eq.~\eqref{beta} but is related by 
\be{beta-}
\beta_-^{-1} =\chi_- T_- 
	= \frac{\chi_-}{\chi_+} \left(\frac{\rho_-}{\rho_+} \right)^{1/4} \beta_+^{-1}  
	.
\ee
In a case, $\beta_-$ could play a role of a temperature with respect to the change of mass $M_{\rm rad}$, which possibility will be discussed in the last section.

Given the temperature $\beta^{-1}$, the heat capacity for fixed volume of the shells is defined by 
\be{CR:shell}
C_V \equiv \Big(\frac{\partial M_{\rm rad}}{\partial \beta^{-1}}\Big)_{r_\pm} =\Big(\frac{\partial M_+}{\partial \beta^{-1}}\Big)_{r_+} - 
\Big(\frac{\partial M_-}{\partial \beta^{-1}}\Big)_{r_\pm} .
\ee
At the present case, the second term vanishes because $M_-$ is held. 
In fact, the fixed volume condition is not transparent because the metric $g_{rr}$ contributes to the volume.
We simply use the terminology to represent that the areal radius of the inner and the outer boundaries do not change.
Direct analytic calculation of the heat capacity is impossible because it requires to solve the corresponding equation of motions analytically, which was solved only numerically in the previous literatures except for a few specific situations. 
However, we find a detour through the variation of entropy in this work.

Introducing scale invariant variables $u$ and $v$ as
\begin{equation} \label{uv:r}
u\equiv \frac{2m(r)}{r}, \qquad v\equiv \frac{dm(r)}{dr} 
= 4\pi r^2 \rho(r) = 4\pi \sigma r^2 T(r)^4,  
\end{equation} 
the variational equation $\delta S_{\rm rad}/\delta m =0$ becomes 
 a first order differential equation for $u$ and $v$,
\begin{equation} \label{dvdu}
\frac{dv}{du} = f(u,v) \equiv \frac{2v(1-2u-2v/3)}{(1-u)(2v-u)}.
\end{equation}
This equation is equivalent to the general relativistic Tolman-Oppenheimer-Volkoff equation of hydrostatic equilibrium for a radiation. 
The allowed range of $(u,v)$ is $u<1$ and $v\geq 0$, where each inequality represents the fact that the spacetimes is static and the energy density of radiations is non-negative, respectively. 
Integrating Eq.~\eqref{dvdu} on the $(u,v)$ plane, solution curves were found in Refs.~\cite{Sorkin:1981wd,Zurek}.
Any solution curve will be parallel to the $u$-axis when it crosses the line
\be{P}
P: ~ 2u + \frac{2v}{3} = 1,
\ee
and is parallel to the $v$-axis when it crosses the line
\be{H}
H: ~ u  = 2v.
\ee
The solution curve eventually converges to the point $\mathcal{R} \equiv (3/7,3/14)$ where the line $P$ crosses $H$.
A specific solution curve $C_\nu$ is characterized by
\be{nu}
\nu \equiv  1-u_H = 1-\frac{2m(r_H)}{r_H},
\ee
the orthogonal distance of $C_\nu$ from the line $u=1$ on the $(u,v)$ plane~\cite{Kim:2016jfh}.
Here the subscript ${}_H$ represents the point where $C_\nu$ crosses $H$, which is the position of the approximate horizon defined by a surface that resembles an apparent horizon~\cite{Kim:2016jfh}.
We quote the name `approximate horizon' from Ref.~\cite{Anastopoulos:2014zqa}.
The value of $\nu$ varies from zero to $\nu_r \approx 0.50735$.
The value $C_0$ and $C_{\nu_r}$ represent solution curves on the verge of the formation of an event horizon and the everywhere regular solution, respectively.
A given solution curve is parameterized by a scale invariant variable 
\be{xi}
\xi \equiv \log \frac{r}{r_H}.
\ee
Therefore, the physically relevant region of $(u,v)$ plane can be equivalently coordinated by using the set $(\nu, \xi)$. 
Now, a specific sphere solution of radiation can be characterized by choosing a boundary point on a curve $C_\nu$ after picking the radius of the boundary $r_+$. 

A given spherical shell of radiation can be denoted by four different numbers, $(\nu, r_H, e^{\xi_+}, e^{\xi_-})$, representing a specific solution curve, the radius of the approximate horizon for the solution curve, and the positions of the inner and the outer boundaries relative to the approximate horizon,  respectively.
The physical parameters at the outer boundary are related with the total mass, the local temperature, and the radius as
\be{r+}
r_+ = r_H e^{\xi_+} , \qquad
u_+\equiv u_\nu(\xi_+) = \frac{2M_+}{r_+}, \qquad 
v_+\equiv v_\nu (\xi_+) = 4\pi r_+^2 \rho(r_+) 
	= 4\pi\sigma r_+^2 T_+^4 , 
\ee
where we put the subscript $\nu$ to $u$ and $v$ to represent the specific solution curve $C_\nu$. 
The physical parameters at the inner boundary are given by
\be{r-}
r_- = r_H e^{\xi_-}, \qquad
u_- \equiv u_\nu(\xi_-) = \frac{2M_-}{r_-},  \qquad 
v_- \equiv v_\nu(\xi_-) = 4\pi r_-^2 \rho(r_-).
\ee
In this work, the value of $u_-$ and $r_-$ are held.
On the other hand, $v_-$ will be determined by tracing in the solution curve $C_\nu$ from the data at the outer boundary.

Even though the static solution of the self-gravitating radiations were studied well, its stability needs additional study.
To achieve this purpose we study its heat capacity.  
In Sec. II, we first derive the relation between the variations of $(u,v)$ and those of $(\nu,\xi)$.  
By using the fact that $(\delta \nu,\delta \xi)$ at the outer boundary is the same as that at the inner boundary if $(u_+,v_+)$ and $(u_-,v_-)$ are on a given solution curve $C_\nu$, we relate the variations of physical parameters at the outer boundary with those at the inner boundary.
In Sec. III, we calculate the heat capacity for fixed volume from the variational equation of entropy.
We show that the general heat capacity is located in the middle of the two extreme forms, that of the regular solution and that of other extreme.  
In Sec. IV, we study various limiting behaviors of the heat capacity.  
We summarize and discuss the results in Sec. V. 
There are three appendices which deal the detailed calculations.

\section{Variations of the scale invariant variables}
The difficulty in calculating the heat capacity of spherical shell of matters lays on the fact that the physical parameters at the inner boundary are dependent on those at the outer boundary, where the exact relation between them requires the knowledge of analytic solutions. 
Rather than searching for an exact analytic solution, we find a variational relation between them.
Because $\delta r_-=0 = \delta M_-$, we have $\delta u_-=0$ leaving only $\delta v_-$ be dependent on the variations at the outer boundary.
We study how to relate the variations at the outer boundary with those at the inner boundary in a general setting. 
To do this, we calculate $\delta \nu$ and $\delta \xi$ corresponding to the variations $(\delta u_+ , \delta v_+)$. 
Then, we use 
(i) The variation $\delta \nu$ is independent of the position of $(u_-,v_-)$ if it is on the same solution curve $C_\nu$ as $(u_+,v_+)$. 
(ii) The variation $\delta \xi = \delta r/r - \delta r_H/r_H$ is also independent of the position of $(u_-,v_-)$ if $r  = r_\pm$ are held.   
(iii) $\nu$ and $\xi$ defines an orthogonal coordinates system which is equivalent to $(u,v)$ physically. 

With these in mind, we find, in the Appendices~\ref{Appendix A} and \ref{Appendix B}, that the variations at the outer boundary are related with those at the inner boundary as
\bea
\delta u_- &=& \frac{f_+f_-}{1+f_+^2}\left( \frac{B_-}{B_+}
 		+\frac{1}{f_+f_-}\frac{2v_--u_-}{2v_+-u_+}  \right)
				\delta u_+  + \frac{f_-}{1+f_+^2}
	\left( -\frac{B_-}{B_+}+\frac{f_+}{f_-}\frac{2v_--u_-}{2v_+-u_+} \right)\delta v_+	 \label{delta u-} , \\
\delta v_-
 &=& \frac{f_+}{1+f_+^2}\left( -\frac{B_-}{B_+}
 			+\frac{f_-}{f_+}\frac{2v_--u_-}{2v_+-u_+}  \right)
				\delta u_+
     + 	 \frac{1}{1+f_+^2}\left( \frac{B_-}{B_+}
 					+f_+f_-\frac{2v_--u_-}{2v_+-u_+} \right)
				\delta v_+	  \label{delta v-} .	
\eea
where $f= F/G$ and $B_\pm \equiv B_\pm(u_\pm, v_\pm)$ is, in Appendix~\ref{Appendix B}, given by 
\be{B:fA}
 B(u,v) = \alpha_\nu  \sqrt{\frac{r_H}{r}} \frac{v^{3/4} \chi G}{
		F^2+ G^2} ; \qquad \alpha_\nu \equiv \frac{2^{3/4}} {3}
		\frac{(7\nu-4)(1-\nu)^{1/4}}{\nu^{1/2}} .
\ee
Here $\nu$ and $r/r_H = e^\xi$ are implicitly dependent on $u$ and $v$ and 
\be{FG}
F \equiv 2v\big( 1-2 u -\frac{2v}{3} \big), \qquad
G \equiv (1-u) (2v-u). 
\ee 
The function $B(u,v)$ goes to zero on $H$ as expected in Eq.~\eqref{fBH}.
It vanishes on $v=0$ and $u=1$ too.
It diverges at the point $\mathcal{R}$.
The proportionality constant $\alpha_\nu$ is negative definite because $\nu$ is restricted to be $0 < \nu \leq \nu_r < 4/7$.

Based on the variational relations~\eqref{delta u-} and \eqref{delta v-}, we obtain, in Appendix~\ref{Appendix B}, 
\bea  \label{dv-:C}
\left(\frac{\partial v_-}{\partial M_+} \right)_{r_\pm,M_-}
 &=&  
  \frac{2}{r_+}\frac{f_-+f_-^{-1}}{f_++ f_+^{-1}} \frac{2v_--u_-}{2v_+-u_+} \left[\frac{1}{f_+} +
  	\frac{2r_+ v_+}{T_+} \C^{-1}
  \right],
\eea
where
\be{C:local}
\C \equiv  \Big(\frac{\partial M_{\rm rad}}{\partial T_+}\Big)_{r_\pm,M_-} .
\ee
The explicit values of $\C$ are obtained at Eq.~\eqref{Clocal:gen} in appendix~\ref{app:C}.

\section{Heat capacity}

The heat capacity~\eqref{CR:shell} can be obtained based on the value of $\C$. 
At the present case, the second term in Eq.~\eqref{CR:shell} vanishes because $M_-$ is held. 
Let us calculate the first term in the right hand side. 
Varying Eq.~\eqref{beta+}, we get
\be{dbeta}
\delta\beta^{-1} =  \frac{M_+T_+}{\chi^2 r_+^2}  \delta r_+  - \frac{T_+}{r_+ \chi} \delta M_+ + \chi \delta T_+ ,
\ee
where we regard $\beta$ as a function of $T_+$, $M_+$, and $r_+$. 
Generally, the three variations $\delta T_+$, $\delta M_+$, and $\delta r_+$ are independent. 
However, if the state inside the inner boundary is invariant under the changes of the physical parameters at the outer boundary, i.e. $\delta r_-=0=\delta M_-$, only two of the three variations will be independent. 
If the size of the shell does not change, $\delta r_+=0$, only one independent variation remains.
In this case, the variations $\delta T_+$ and $\delta M_+$ must be related.
Dividing Eq.~\eqref{dbeta} by $\delta M_+$ we find that the heat capacity~\eqref{CR:shell} is related with $\C$ by
\be{dM1:dbeta}
\frac{1}{\beta^{-1}C_V} =  
	\frac{1}{T_+\C } -\frac1{r_+ \chi^2} . 
\ee
Note that the heat capacity $C_V$ is positive when 
$$
 0 < \C < \frac{r_+ \chi^2}{T_+} .
$$
Therefore, the positivity of the heat capacity is not always guaranteed by the positivity of $\C$.

Inserting the value of $\C$ in Eq.~\eqref{Clocal:gen} to Eq.~\eqref{dM1:dbeta}, the heat capacity for a shell of radiations is given by
\bea \label{cv:gen}
C_V &=& \frac{r_+\chi^2}{\beta^{-1}} \frac{1- \A}{
	\chi^2f_+/(2v_+) -1 + (\chi^2/(2v_+f_+) +1) \A} , \qquad \mathfrak{A} \equiv \sqrt{\frac{r_-}{r_+}} \frac{A_+}{A_-},
\eea
where $A_\pm \equiv A(u_\pm, v_\pm)$ with 
\bea \label{Apm}
(2v-u) A(u,v) \equiv
   	\frac{v^{3/4} }{\chi }	
		\frac{f}{(2v -u)(1+f^2)} 
	=  \frac{\chi v^{3/4} F}{F^2 + G^2}.
\eea   
Note that the function $(2v-u) A(u,v)$ is a regular function over the whole range of physical interest other than the point $\mathcal{R}$, where $\mathcal{R}$ corresponds to the asymptotic infinity $r\to \infty$ of all solution curves. 
It vanishes on the lines $P$ and $v=0$.
The function $A(u,v)$ is positive definite in the region with $u \to -\infty$ and changes signature when a solution curve crosses the lines $P$ and $H$.

In the limit $\A \to 0$, the heat capacity reproduces that of the regular sphere given in Ref.~\cite{Pavon1988}:
\be{cR1:reg}
C_{V}^{\rm reg} =\Big(\frac{\partial M_+}{\partial \beta^{-1}}\Big)_{r_+} 
	= - \frac{r_+ \chi^2}{\beta^{-1}} \frac{2v_+-u_+}{ 8v_+/3-1+ u_+}.
\ee
The heat capacity changes sign on the line $H$ and is singular on the line
\be{Q}
Q: \frac{8v_+}{3} + u_+ =1.
\ee

\begin{figure}[th]
\begin{center}
\begin{tabular}{c}
\includegraphics[width=1.\linewidth, scale=2, trim = 15mm 224mm 15mm 18mm,clip]{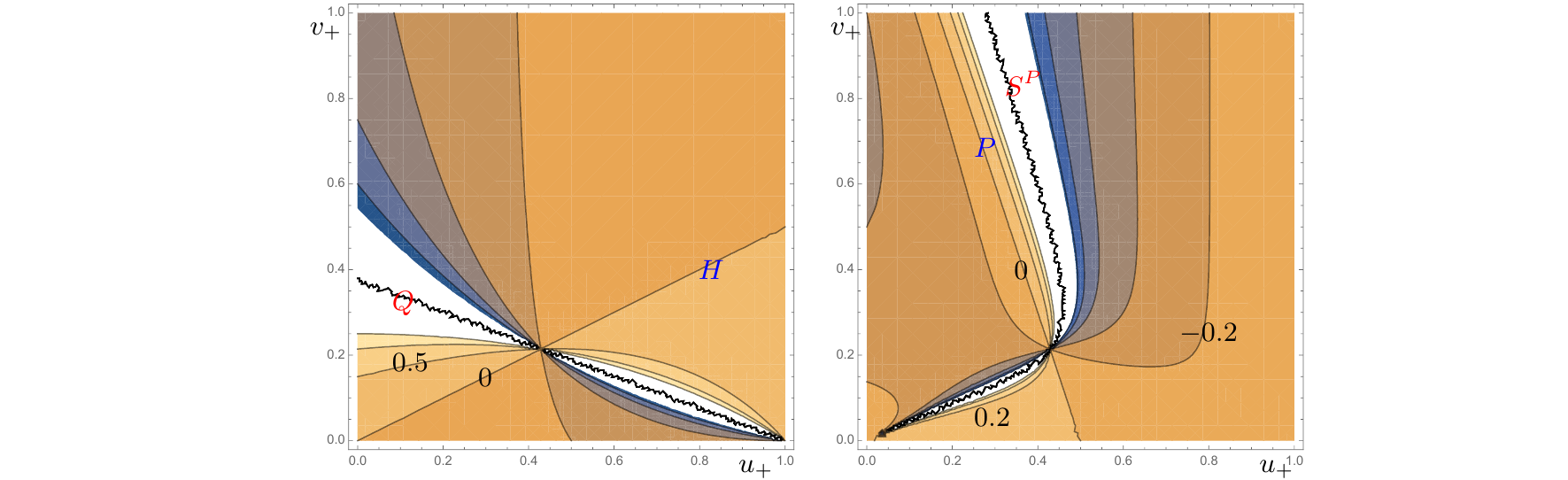}
\end{tabular}
\end{center}
\caption{ Heat capacities, $\beta^{-1} C_V/r_+$, for the limits of $\A \to 0$ (L) and $\A \to \infty$ (R).   
}
\label{fig:CVlimits}
\end{figure}
On the opposite limit $\A \to \infty$, we have 
\be{CV:infty}
C_V^P \equiv \lim_{\A \to \infty} C_V  = -\frac{r_+ \chi^2}{\beta^{-1}} 
	\frac{4v^2 (1-2u_+ - 2v_+/3)}{(1-u_+)^2(2v_+ - u_+) + 4v_+^2(1- 2u_+ - 2v_+/3)}.
\ee
As shown the the right panel of Fig.~\ref{fig:CVlimits}, the heat capacity in this limit vanishes on the line $P$ and is singular on the curve 
\be{CC}
S^P: (1-u_+)^2(2v_+ - u_+) + 4v_+^2(1- 2u_+ - 2v_+/3) =0.
\ee
The curve passes the point $\mathcal{R}$.
For  $v_+\ll 1$, it overlaps with the line $H$ and, for $v_+\gg 1$, approaches the line
\be{CC:large}
1-u_+ = \gamma v_+, \qquad \gamma = \frac23 \left( -1 - \frac{2^{2/3}}{(4+ 3\sqrt{2})^{1/3}}
	+ (8 + 6\sqrt{2})^{1/3} \right) \approx 0.5062.
\ee
These behaviors are manifest in the right panel of Fig.~\ref{fig:CVlimits}.
Note that the form of the heat capacity are completely different from that of the regular solution.  
As for a regular solution, $C_V^{\rm reg}$ is singular on the line $Q$ and changes sign on $H$. 
On the other hand, $C_V^P$ is singular on $S^P$ and vanishes on the line $P$.
This implies that the excision of the central conical singularity plays an important role in the thermodynamics of the system.

\begin{figure}[t]
\begin{center}
\begin{tabular}{c}
\includegraphics[width=1.\linewidth, scale=2, trim = 15mm 224mm 15mm 18mm,clip]{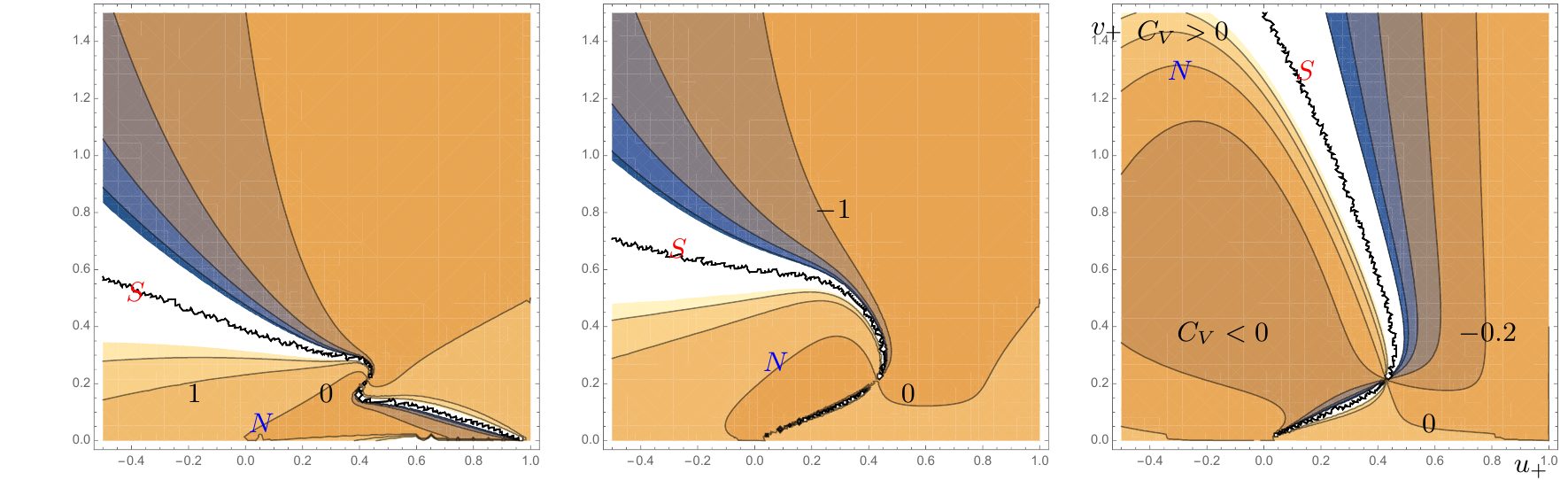}
\end{tabular}
\end{center}
\caption{ $\beta_+^{-1}C_V/r_+$ on the plane $(u_+,v_+)$. 
The heat capacities for $A_-(r_+/r_-)^{1/2} =20, 1,$ and $1/20$, respectively from the left.  
}
\label{fig:cv}
\end{figure}
Writing the heat capacity~\eqref{cv:gen} explicitly in terms of $(u_+, v_+)$, we get
\bea \label{cv:gen2}
C_V 
 &=&  \frac{r_+\chi^2}{\beta^{-1}} \frac{ (2v_+-u_+) (F_+^2+ G_+^2)  
 	- \frac{v_+^{3/4} \chi_+}{A_- (r_+/r_-)^{1/2}} F_+}
 	{\big(1-u_+ - \frac{8v_+}{3}\big) (F_+^2+ G_+^2) + \frac{v_+^{-1/4} \chi_+}{2A_- (r_+/r_-)^{1/2}} 
		\big( 2v_+F_+ + \chi_+^2 G_+ \big)}.
\eea
Note that the information at the inner boundary come into with the combination of $A_-r_-^{-1/2}$.
The denominator of Eq.~\eqref{cv:gen2} vanishes on the curve given by
\be{CV:D}
S:~\Big(1-u_+ - \frac{8v_+}{3}\Big) (F_+^2 + G_+^2) 
	+ \frac{v_+^{-1/4} \chi_+}{2A_- (r_+/r_-)^{1/2}} 
		\Big(2v_+F_+ + \chi_+^2 G_+ \Big) =0 . 
\ee
On this curve, the heat capacity is singular.
The curve passes $\mathcal{R}$ along the curve $S^P$ because $F_+ \to 0$ and $G_+ \to 0$ at $\mathcal{R}$ leaving the last term in Eq.~\eqref{CV:D} as the first nontrivial corrections. 
Equation~\eqref{CV:D} indicates that the singular curve $S$ must be around the line $Q$ and the curve $S^P$ when $A_-(r_+/r_-)^{1/2} \gg 1$ and $A_-(r_+/r_-)^{1/2} \ll 1$, respectively. 
These behaviors are manifest in the left and right panels of Fig.~\ref{fig:cv}, respectively.
As $F_+$ or $G_+$ are larger, i.e., $\chi$ or $v_+$ increases, the singular curve gradually approaches the line $Q$.  
Combining the pictures in Figs.~\ref{fig:CVlimits} and \ref{fig:cv}, one may find that the singular curve gradually change from the line $Q$ to the curve $S^P$ as $A_-(r_+/r_-)^{1/2}$ decreases.

The numerator of Eq.~\eqref{cv:gen2} vanishes on the curve 
\be{CV:C0}
N: (2v_+ -u_+) (F_+^2 + G_+^2) - \frac{v_+^{7/4} \chi}{A_-(r_+/r_-)^{1/2}} 
	\Big( 1- 2u_+ - \frac{2v_+}{3} \Big) =0 .
\ee
The heat capacity changes sign on $N$.
The curve passes the point $\mathcal{R}$ along the line $P$ because $F_+ \to 0$ and $G_+ \to 0$ at $\mathcal{R}$ leaving the last term in Eq.~\eqref{CV:C0} as the first nontrivial corrections.  
Equation~\eqref{CV:C0} indicates that the curve $N$ must be located around the line $H$ for $A_-(r_+/r_-)^{1/2} \gg 1$ and around $P$ for $A_-(r_+/r_-)^{1/2} \ll 1$, respectively. 
These behaviors are manifest in the left and right panels of Fig.~\ref{fig:cv}, respectively. 
Combining the pictures in Figs.~\ref{fig:CVlimits} and~\ref{fig:cv}, one may find the the  curve $N$ gradually changes from the line $H$ to the line $P $ as $A_-(r_+/r_-)^{1/2}$ decreases.
When viewed from the clockwise direction centered at the point $\mathcal{R}$, the heat capacity takes positive values from $N$ to $S$ and negative values elsewhere.

\section{Various limits}
Because the functional forms of the heat capacity are complicated, we present various physically interesting limits to improve our understanding on the system. 

\subsection{Thin shell limit} \label{sec:IVA}
We first consider the thin shell limit, $\delta r \equiv r_+ -r_- \ll r_+ \approx r$, $\delta u \equiv u_+-u_- \ll u_+$, and $ \delta v \equiv v_+ - v_- \ll v_+$. 
By using 
$$
\delta r  =  r \delta \xi = \frac{r \delta u}{2v - u}, \qquad 
\delta v = f(u,v) \delta u,
$$
in this limit, the heat capacity takes the form, 
\bea
\frac{\chi^{-1} C_V}{\delta r} &\approx& \frac{v}{T(f+ f^{-1})} \left[1 - 2r \frac{\delta \log A}{\delta r}   \right]\nn \\
   &\approx& \frac{2 v (2v-u)FG}{3T(F^2+G^2)} 
   \left[\frac{4v^2 - 4v -u^2 +28uv/3}{(1-u) (2 v-u)^2}
   	-\frac{2}{1-2
   u-2 v/3}  \right. \nn\\
  &&+ \left.  \frac23 \frac{-16v^3+4 (7-13 u) v^2+2 (u-1) (5 u-3) v-9 u(u-1) (2 u-1)}
  			{ F^2 + G^2}
   ~  \right] .
\eea
Interestingly, the heat capacity is regular over all physically allowed values of $(u,v) \neq \mathcal{R}$. 
Even though it can take both signatures, its value change smoothly. 

\subsection{$r_- \ll r_H$ approximation} \label{sec:IVB}
Let us next consider the `almost sphere' case which excises only the central singularity by using the limit $r_- \to 0$.
We are interested in a solution having a conical singularity at the center, i.e. $\nu \neq \nu_r$.
Solving Eqs.~\eqref{uv:r} and \eqref{dvdu} around the center (or simply quoting results in Ref.~\cite{Kim:2016jfh}), one gets approximately
\be{uv:0}
m(r_-) = -\frac{\mu_0 r_H}{2}+ \frac{\kappa r_H}{10} \left(\frac{r_-}{r_H}\right)^5 , \qquad
u_- = -\frac{r_H\mu_0}{r_-} , 
\qquad
v_- = \frac{\kappa}{2} \left(\frac{r_-}{r_H}\right)^4 = \frac{\kappa\mu_0^4}{2}  u_-^{-4}.
\ee
Note that there is a central conical singularity with negative mass at $r=0$ unless it is excised. 
By using the results in Eq.~\eqref{uv:0}, we get
$$
A_-  =  \frac{v_-^{3/4}\sqrt{1-u_-}}
	{2v_--u_-}
	\frac{F_-}{F_-^2+G_-^2} 
= 4 \frac{v_-^{7/4}}{|u_-|^{7/2}}
=  4 \left(\frac{v_-}{u_-^2} \right)^{7/4} = 4 \left(\frac{\kappa \mu_0^4}{2u_-^6} \right)^{7/4}  .
$$
Therefore, $A_- (r_+/r_-)^{1/2}  \ll 1$ in the limit. 
Now, the heat capacity takes the form in Eq.~\eqref{CV:infty}.
Its behaviors are shown on the right panel of Fig.~\ref{fig:CVlimits}.

Let us observe the case that both boundaries are located around the center, $r_-\ll r_+ \ll r_H$.
The entropy of the system is given by 
\be{Srad:sm r}
S_{\rm rad} = S_+ - S_- = \frac{r_H^{3/2} }{6} 
 	\left(\frac{8\pi \sigma }{\kappa \mu_0^2 }\right)^{1/4}
		\frac{r_+-r_-}{r_H} +\cdots.
\ee
where 
$$
S_\pm = \frac{r_H^{3/2} }{3} \frac{e^{3\xi_\pm/2}}{\sqrt{1-u_\pm}}
	\left(\frac{4\pi \sigma}{v_\pm}\right)^{1/4} \Big(
		\frac{2v_\pm}{3} + u_\pm\Big) 
 \approx \frac{r_H^{3/2} }{3} 
 	\left(\frac{8\pi \sigma \mu_0^2}{\kappa }\right)^{1/4}
		\Big(-1 + \frac{1}{2\mu_0} \frac{r_\pm}{r_H} +\cdots \Big).
$$
Note that $S_\pm$ has a non-vanishing negative constant contribution in the $r\to 0$ limit.  
The heat capacity of the system is independent of the information at the inner boundary and takes negative value,
$$
C_V \approx - \frac{2\kappa^2r_H}{\mu_0} \left(\frac{r_+}{r_H}\right)^{10} + \cdots .
$$
Therefore, the system must be unstable under perturbations.
The heat capacity becomes positive after the solution curve passes the line $P$, where approximation $r_+ \ll r_H$ does not hold any more.

\subsection{Near approximate horizon case} \label{sec:IVC}
We next consider the case that $r_-$ is located around the approximate horizon.
We assume that the approximate horizon is about to form an event horizon, $\nu \sim 0$.  
A special case is that the inner boundary is located exactly at the approximate horizon. 
Then, the heat capacity is given by that of the regular solution as discussed in the paragraph just after Eq.~\eqref{Clocal:gen}.

First, let us consider the case that both boundaries are located around the approximate horizon, $1-u_\pm \ll 1$ and $\varepsilon^2\ll  v_+< v_- \ll \varepsilon^{-2/3}$, where $ \varepsilon = 9\nu/16$ is a small expansion parameter.
In this region,\footnote{This region corresponds to $[\mathfrak{R,S}]$ in Ref.~\cite{Kim:2016jfh}.} the solution curve $C_\nu$ satisfies~\cite{Kim:2016jfh}  
\be{1-u}
1-u \approx  \varepsilon \frac{ (2v/3+1)^2}{\sqrt{2v}} + O(\varepsilon^2).
\ee
The radius is given by
\be{rv:Q}
r =r_H e^\xi ; \qquad \xi  =\frac{\varepsilon }{\sqrt{2v}}\left(1 -\frac{v}{6} \right)- \frac{11\varepsilon}{12},
\ee
where we choose $\xi =0$ at $H$.
Note that $r$ changes only a bit for a large change of $v$ in this region. 
This gives, by using $r_+ \simeq r_- \simeq r_H$ and $u \simeq 1$, 
$$
\frac{(2v-u) A}{\sqrt{r}}\approx
-2^{-5/4} \sqrt{\frac{\varepsilon}{r_H v}} 
~ \Rightarrow ~ \A = \sqrt{\frac{v_-}{v_+}} \frac{2v_--1}{2v_+ -1}. 
$$
Therefore, the  heat capacity becomes
\be{HC:AH}
\chi^{-1}C_V \approx 
- \frac{\varepsilon \, r_+}{\sqrt{2}v_+T_+} \Big(\frac{2v_+}3 +1\Big)
	(\sqrt{v_+} - \sqrt{v_-}) [ 2(v_+ + v_- + \sqrt{v_+ v_-} ) -1] .
\ee
Because $v_+ < v_-$, the sign of the heat capacity is determined by the sign of $2(v_+ + v_- + \sqrt{v_+ v_-} ) -1$.
For $v_-\geq 1/2$, the heat capacity is positive definite. 
The heat capacity becomes negative only if $v_- < 1/2$ and 
$v_+ < ( \sqrt{2-3v_-}- \sqrt{v_-})/2$. 

We next consider the case that the both boundaries are located in the region outside the horizon satisfying $v_\pm \ll 1$ and $\varepsilon^{2/3} \ll u_+< u_- < 1$.\footnote{This corresponds to the region  $[\mathfrak{S},\mathfrak{S}' ]$ in Ref.~\cite{Kim:2016jfh}.}
In this case, $r$, $u$ and $v$ are related by
\be{SS'}
v = \frac{\varepsilon^2}{2u^2 (1-u)^2}, \qquad 
r \approx \frac{r_H}{u} \left(1-\frac{11\varepsilon}{12} \right)+\cdots .
\ee
Then, the function $A$ is given by
$$
\frac{A}{\sqrt{r}} 
	\approx \frac{\varepsilon^{7/2}}{2^{3/4} r_H^{1/2}} \frac{2u-1 }{u^{6} \chi^{10}}.
$$
Putting this to Eq.~\eqref{cv:gen}, the heat capacity becomes
\be{HC:AH2V}
C_V =  \frac{r_H \chi_+^2}{\beta^{-1} u_+} \frac{u_-^6 \chi_-^{10}}{2u_--1}
	\left( \frac{2u_+ -1}{u_+^6 \chi_+^{10} } -  
		\frac{2u_- -1}{u_-^6 \chi_-^{10} }
	\right).
\ee
Because $(2u-1)/(\chi^{10}u^{6}$) is a monotonically increasing function of $u$ and $u_+< u_-$, the terms inside the parenthesis is negative definite.
Therefore, 
$C_V$ is positive definite because $u_- < 1/2$ in this region. 

Finally, we consider the case that the inner and the outer boundaries are located around the approximate horizon and outside of the approximate horizon, respectively.
The heat capacity is given by
\be{HC:AH3V}
C_V = \frac{ \varepsilon r_H }{\beta^{-1}} 
	\frac{1}{\varepsilon u_+
	+ \sqrt{2v_-} (2v_--u_-)}
	\approx
 \frac{ \varepsilon r_H}{\beta^{-1}} \frac{1}{\sqrt{2v_-}(2v_--u_-)} , 
\ee
where the last equality is valid unless $2v_- = u_-$.
Usually, the value of heat capacity is of $O(\varepsilon)$. 
If the inner boundary is located on $H$, the heat capacity suddenly jumps to $O(1)$, which value is the same as that of the regular solution in Eq.~\eqref{cR1:reg} as expected just after Eq.~\eqref{cv:gen}.

\section{Summary and discussions}
In this work, we have studied analytically the heat capacity of a static spherically symmetric self-gravitating radiations in the context of general relativity.
To avoid ambiguity due to the central conical singularity, we excise the central region and introduce an inner boundary at $r_-$. 
Then, the system inside the inner boundary is assumed to be filled with matters of spherically symmetric distribution with its total mass $M_-$ being held. 
Therefore, the radiations are confined inside a shell bounded by two stiff boundaries at $r_+$ and $r_-$.
The distribution of the radiations and the geometry are described by putting the boundary values on a solution curve $C_\nu$ on a two dimensional plane $(u = 2m(r)/r, v = 4\pi r^2 \rho )$ of scale invariant variables, which curve can be found by solving the TOV equation.

We first derived how to relate the variations at the outer boundary with those at the inner boundary. 
At the outer boundary, there are three independent variables, $r_+$, $M_+$, and $\rho_+$.
To obtain the heat capacity of the radiations for fixed volume, we have assumed that the location of the inner and the outer boundaries are held.
Then at the outer boundary, there remains two independent variables which can be varied.   
Because $M_-$ and $r_-$ are held, only $v_-$ can be varied at the inner boundary. 
The variation $\delta v_-$ will induce the variations $\delta u_+$ and $\delta v_+$ at the outer boundary through the TOV equation. 
Because there are only one variation at the inner boundary, the variations at the outer boundary should be related and the relation shows up as a heat capacity.
To get an analytic form for the heat capacity, we additionally use the thermodynamic identity $\delta S_{\rm rad} = \beta \delta M_{\rm rad}$, which is derived from the variation of the entropy formulae.

Let us display a few interesting results. %
There are two limiting forms for the heat capacity.
i) When the inner boundary is located at the approximate horizon, the heat capacity of the shell of the radiations are the same as that of the self-gravitating sphere of regular solution. 
ii) When the inner boundary is located on the line $P: 2u_-+ 2v_-/3=1$, the heat capacity shows other limiting form much different from that of the regular one. 
As the outer boundary changes, a heat capacity may take singular or null values at a specific point on the $(u_+,v_+)$ plane. 
The singular curve $S$ changes from $Q: 8v_+/3+u_+=1$ to $S^P: (1-u_+)^2(2v_+-u_+)+ 4v_+^2(1-2u_+- 2v_+/3) =0 $.
On the other hand, the null curve $N$ changes from $H$ to $P$. 
When viewed from the clockwise direction centered on $\mathcal{R}$, the heat capacity is positive definite from $N$ to $S$.
For the case of the zero size limit of the inner boundary, $r_- \to 0$, it was shown that the heat capacity does not go to the form of the regular solution. 
Rather, they approaches the opposite limit ii) unless the solution curve is that of the regular one. 
Finally, we have obtained the heat capacity for the case that both boundaries are located around the approximate horizon.  
We find that there are no singularity of heat capacity contrary to the general case.

An interesting topic is that the possibility to define a new heat capacity such as $\C$. 
The heat capacity provides an important criterion for determining stable equilibrium.
For the case of the heat capacity $C_V$, the concavity of the entropy is directly related with the positivity of $C_V$ because $\delta^2 S/\delta M_+^2 = -\beta^2 C_V^{-1}$. 
However, for the case of $\C$, the concavity may not be directly related with the positivity of $\C$.
This is because the energy inside $r_+$ with respect to a local observer is not given by $M_+$. 
In this sense, $\C$ cannot play a role discriminating the concavity of the entropy.  

An interesting discovery is that the variation of the entropy of the system of the radiations in a spherical shell is related not simply with the variation of the radiation's mass but also with the variation of the mass inside the inner boundary. 
Once the radiations satisfy the equation of motions in Eq.~\eqref{dvdu}, the variational relation~\eqref{dSrad} of the entropy is given with the variations at the boundaries by
\be{dSrad:dM}
\delta S_{\rm rad} = \beta_+ \delta M_+ -\beta_- \delta M_- = \beta_+ \delta M_{\rm rad} +
	(\beta_+ -\beta_-) \delta M_-
= (\beta_+-\beta_-)\delta M_+ +\beta_- \delta M_{\rm rad}
.
\ee
The present law is different from the ordinary thermodynamic first law in the sense that 
the entropy variation is dependent not only to $\delta M_{\rm rad}$ but also to $\delta M_-$. 
When $\delta M_-=0$ or $\beta_+= \beta_-$, one can identify $\beta^{-1} \equiv \beta_+^{-1}$ as the temperature measured in the asymptotic region.
On the other hand, when $\delta M_+ =0$ with $\delta M_- \neq 0$, i.e. the outer boundary isolates the system from the outside thermodynamically and the heats flow through the inner boundary, $\beta_-^{-1}$ plays the role of a temperature in the sense that $\delta S_{\rm rad} = \beta_- \delta M_{\rm rad}$.  
However, $\beta_-^{-1}$ is different from the asymptotic temperature $\beta^{-1}$ and is not directly related with the local temperature $T_-$ by the Tolmann formula unless 
$ \rho_+ = (\chi_-^4/\chi_+^4) \rho_-$.  
In this sense, the variational relation~\eqref{dSrad:dM} appears to admit two different legitimate temperatures depending on physical situations. 
Mathematically, the origin of this dual temperatures is $g_{tt}(r_-) g_{rr}(r_-) \neq -1$.
A physical explanation for this difference is that the change of the mass of the radiation, $\delta M_{\rm rad}$, through the inner boundary must accompany with the change of the mass inside the inner boundary, which modifies not only the thermodynamic situation but also the gravity of the shell through the Birkhoff's theorem. 
On the other hand, the change of mass outside of $r_+$ does not affect the gravity inside directly.
One may define a heat capacity based on $\beta_-$ too, which may raise a new instability problem of the system. 
Physical implication of $\beta_-$ needs further studies in the future research.

\section*{Acknowledgment}
This work was supported by the National Research Foundation of Korea grants funded by the Korea government NRF-2017R1A2B4008513.
The author thanks to APCTP.

\appendix

\section{Generic variations} \label{Appendix A}
Let us consider the variation $\delta \xi$ and $\delta \nu$, which represent the variations parallel to and orthogonal to the solution curve $C_\nu$, respectively.
The two variations are orthogonal to each other and define most general changes of boundary points, $(u,v) \equiv (u_\pm, v_\pm)$, on $C_\nu$.
An important point here is that $\nu$ is independent of the choice of the boundary point on $C_\nu$ by definition and the variation of $\xi_\pm = \log (r_\pm /r_H)$ is required to be dependent only on the change of $r_H$ because $r_\pm$ are held. 
Therefore, we have $\delta \xi_+ = \delta r_H/r_H = \delta \xi_-$.
In this subsection, we omit the subscript $\pm$ for notational simplicity. 

From Eqs.~\eqref{uv:r} and \eqref{xi}, we get $(\partial u/\partial \xi) = 2v-u$. 
By using Eq.~\eqref{dvdu}, the tangent along $C_\nu$ is 
\be{partial xi}
\frac{\partial}{\partial\xi}   
	= (2v-u)\frac{\partial}{\partial u} +(2v-u)f(u,v) \frac{\partial}{\partial v} ,
\ee
where $f(u,v) = (dv/du)_\nu$ along the solution curve given in Eq.~\eqref{dvdu}.
On the other hand, the derivative orthogonal to $C_\nu$ can be written as 
\be{partial nu}
\frac{\partial}{\partial\nu} 
 = -B(u,v) f(u,v)\frac{\partial}{\partial u}
    +B(u,v) \frac{\partial}{\partial v} , 
\ee
where we use $\partial/\partial \xi \perp \partial/\partial \nu$ and $B$ is a local function of $(u,v)$ defined by
\be{B} 
    B(u,v) \equiv \left(\frac{ \partial v}{\partial\nu}\right)_{\xi}.
\ee
The explicit functional form of $B$ will be determined at Eq.~\eqref{B:fA} in Appendix~\ref{Appendix B} from consistency. 
From Eqs.~\eqref{partial xi} and \eqref{partial nu}, we determine $\delta u$ and $\delta v$ in terms of $\delta \nu$ and $\delta \xi$ as,
\bea 
\delta u 
	=(2v-u) \delta \xi -B f \delta \nu , \qquad 
\delta v  
	= (2v-u) f \delta \xi + B \delta \nu. \label{delta v}
\eea
Inverting Eq.~\eqref{delta v}, 
the variation $\delta \nu$ and $\delta \xi$ are given by 
\bea
\delta \nu 
	= \frac{-f \delta u + \delta v}{B(1+f^2)} , \qquad 
\delta \xi 
	= \frac{\delta u + f \delta v}{(2v-u)(1+f^2)}  .\label{delta xi}
\eea
Let us see the results at the point $r= r_H$ where $u_H= 2v_H$.  
At this point, $\delta \xi$ and $\delta \nu$ are parallel to $\delta v$ and $\delta u$, respectively.
Therefore,  $(\delta u/\delta \xi)_{r=r_H} = 0$, $(\delta v/ \delta \nu)_{r=r_H} = 0 $. 
In addition, from Eq.~\eqref{delta v},
\be{deltas}
\delta v_H = \left[(2v - u) f(u,v) \right]_{r\to r_H} \delta \xi = -\frac{2(14 v_H/3-1)}{1- 2v_H} v_H \delta \xi, 
\qquad
\delta u_H  = -\left[\lim_{r\to r_H} B(u,v) f(u,v)\right] \delta \nu  .
\ee
From the first equation, one notes that $\delta v_H$ diverges as $v_H \to 1/2$, which corresponds to the limit of forming an event horizon. 
The second equation, by using Eq.~\eqref{nu}, determines the normalization of $B$ to be 
\be{fBH}
\lim_{r\to r_H} f(u,v)B(u,v) = 1.
\ee 
In Appendix~\ref{Appendix B}, we finalize the function $B$ from this normalization condition.
Because $f(u,v)$ diverges on $H$, the value of $B$ vanishes there.

\section{Variations at the inner and the outer boundaries} \label{Appendix B}

By using the fact that $\delta \nu$ and $\delta \xi$ are independent of the position on a given solution curve $C_\nu$, we relate the variations at the outer boundary with those at the inner boundary.
From Eqs.~\eqref{delta v} and \eqref{delta xi}, the variation of $u_-$ can be written by the variations at the outer boundary as 
\bea
\delta u_- &=&   -B_-f_- \delta \nu 
	+ (2v_--u_-) \delta \xi 
\nn \\
 &=& \frac{f_+f_-}{1+f_+^2}\left( \frac{B_-}{B_+}
 		+\frac{1}{f_+f_-}\frac{2v_--u_-}{2v_+-u_+}  \right)
				\delta u_+  + \frac{f_-}{1+f_+^2}
	\left( -\frac{B_-}{B_+}+\frac{f_+}{f_-}\frac{2v_--u_-}{2v_+-u_+} \right)\delta v_+	 \label{delta u-2} ,
\eea
where $B_\pm$ and $f_\pm$ stand for $B(u_\pm, v_\pm)$ and $f(u_\pm, v_\pm)$, respectively. 
Using Eq.~\eqref{C:local} after dividing Eq.~\eqref{delta u-2} by $\delta M_+$, we get
\bea  \label{du=0}
 \left(f_+- \frac{2r_+v_+}{T_+} \C^{-1}\right)\frac{B_-}{B_+} =-
 	\left( \frac{1}{f_+} + \frac{2r_+v_+}{ T_+} \C^{-1} \right)\frac{f_+}{f_-} \frac{2v_--u_-}{2v_+-u_+},
\eea
where we use $\left(\partial u_-/\partial M_+\right)_{r_\pm,M_-}=0$ because $r_-$ and $M_-$ are held.
In a similar manner, the variation of $v_-$ can be written by means of the variations at the outer boundary as
\bea
\delta v_- &=&  B_- \delta \nu 
	+ (2v_- - u_-)f_- \delta \xi 
		\nn \\
 &=& \frac{f_+}{1+f_+^2}\left( -\frac{B_-}{B_+}
 					+\frac{(2v_--u_-)f_-}{(2v_+-u_+)f_+}  \right)
				\delta u_+
     + 	 \frac{1}{1+f_+^2}\left( \frac{B_-}{B_+}
 					+\frac{(2v_--u_-) f_-f_+}{2v_+-u_+} \right)
				\delta v_+	  \label{delta v-2} .
\eea
Using Eq.~\eqref{C:local} after dividing Eq.~\eqref{delta v-2} by $\delta M_+$, we get
\bea \label{dv/dM+}
\left(\frac{\partial v_-}{\partial M_+} \right)_{r_\pm,M_-}
&=& 
  \frac{2}{r_+}\frac{1}{1+ f_+^2}\left[- \left( f_+- \frac{2r_+ v_+}{ T_+} 	
  		\C^{-1}  \right) \frac{B_-}{B_+} 
  + f_+f_- \frac{2v_--u_-}{2v_+ - u_+} 
  	\left(\frac{1}{f_+} + \frac{2r_+ v_+}{T_+} \C^{-1} \right)\right]. 
\eea
Putting Eq.~\eqref{du=0} to Eq.~\eqref{dv/dM+}, one gets 
\bea  \label{dv-:C2}
\left(\frac{\partial v_-}{\partial M_+} \right)_{r_\pm,M_-}
 &=&  
  \frac{2}{r_+}\frac{f_-+f_-^{-1}}{f_++ f_+^{-1}} \frac{2v_--u_-}{2v_+-u_+} \left[\frac{1}{f_+} +
  	\frac{2r_+ v_+}{T_+} \C^{-1}
  \right].
\eea
Once we get $\C$ explicitly we can obtain the function $B$ from Eq.~\eqref{du=0} in addition to the relation between the outer boundary and the inner boundary through Eq.~\eqref{dv-:C2}. 
The explicit form of the function $B$ will be calculated in Eq.~\eqref{B:fA} in a subsequent appendix.

\section{Calculation of Heat Capacity} \label{app:C}

Direct calculation of the heat capacity needs to solve the equation of motion~\eqref{dvdu} from $r_-$ to $r_+$, which is impossible analytically. 
On the other hand, the entropy has an exact analytic expression. 
Fortunately, the integration in Eq.~\eqref{entropy} can be executed to give an analytic form for the entropy of the radiation of the shell~\cite{Sorkin:1981wd,Chavanis:2007kn}, 
\be{S:rad}
S_{\rm rad}  \equiv S_+- S_- ,\qquad
S_\pm(u_\pm,v_\pm,r_\pm)=\frac{r_\pm^{3/2}}{3 \chi_\pm} 
	\left(\frac{4\pi \sigma }{v_\pm }\right)^{1/4}
		\Big(\frac{2v_\pm}3 + u_\pm\Big)    
= \frac{r_\pm \beta_\pm }{3 } \Big(\frac{2v_\pm}3 + u_\pm\Big).
\ee  
Remember that $S_\pm$ does not represent the entropy of the objects inside $r_\pm$ unless the contribution from the central conical singularity vanishes.
For later convenience, we put the derivative of $S_{\pm}$ as
\be{dS:duv}
\beta_\pm^{-1} d S_\pm = \frac{1}{2}(\frac{2v_\pm}3 +u_\pm)\,dr_\pm
	+ \frac{r_\pm}{6}\frac{2-u_\pm+ 2v_\pm/3}{1-u_\pm} du_\pm 
	+ \frac{r_\pm}{12}\frac{2v_\pm-u_\pm}{v_\pm} dv_\pm.
\ee
If we consider on-shell variations [$du$ and $dv$ are related by Eq.~\eqref{dvdu}], we get the first law of thermodynamics, $dM_\pm = \beta_\pm^{-1} dS_\pm - p_\pm (4\pi r_\pm^2) dr_\pm$ from this equation even though $S_\pm$ does not represent the entropy of the corresponding system inside.

Therefore, it would be better to use Eq.~\eqref{dSrad:dM} to obtain the heat capacity. 
We assume that the radiation is thermodynamically isolated from the matters at $r<r_-$. 
Therefore, the mass inside the inner boundary must be independent of the thermodynamic changes of the radiations, which requires $\delta M_- =0$.  
Then, Eq.~\eqref{dSrad:dM} becomes
\bea \label{dS/dm+}
0 =\left(\frac{\partial S_{\rm rad}}{\partial M_{\rm rad}}\right)_{r_\pm, M_-} -\beta
= \left(\frac{\partial S_+}{\partial M_+}\right)_{r_+} - \beta -\left(\frac{\partial S_-}{\partial M_+}\right)_{r_\pm,M_-} , 
\eea
where $S_{\rm rad}$ is given in Eq.~\eqref{S:rad}.
Because $r_\pm$ and $M_-$ are held, $S_+$ and $S_-$ are local functions of $(u_+,v_+)$ and $v_-$, respectively.

Before dealing complex general cases, let us review how the heat capacity for a self-gravitating radiation sphere with regular center was calculated in Ref.~\cite{Pavon1988} by choosing $\nu=\nu_r$ and $r_-=0$. 
Because $S_-=0$, the last term in the right-hand side of Eq.~\eqref{dS/dm+} vanishes. 
Noting $r_+$ is held, by using Eqs.~\eqref{r+} and \eqref{dS:duv},  the right-hand side of Eq.~\eqref{dS/dm+} becomes 
\bea \label{dS+dm+}
\left(\frac{\partial S_{+}}{\partial M_{+}}\right)_{r_+} -\beta  
&=& \frac{2}{r_+}
	\left(\frac{\partial S_+}{\partial u_+}\right)_{r_+,v_+}
	+\frac{4v_+}{T_+} 
		\left(\frac{\partial T_+}{\partial M_+}\right)_{r_+} \,
	\left(\frac{\partial S_+}{\partial v_+}\right)_{r_+,u_+} -\beta
\nn \\
&=& \frac{\beta r_+(2v_+-u_+)}{12 v_+} \left[ - \frac{2f_+}{r_+} + 
	\frac{4v_+}{T_+} \left(\frac{\partial T_+}{\partial M_+}\right)_{r_+} \right] .
\eea
For a regular solution, there remains only one free degree of freedom in the physical parameters at the outer boundary because the size $r_+$ is held.
Therefore, the variations $\delta u_+$ and $\delta v_+$ must be dependent on each other, which relation determines $\C$. 
Now, $\C$ for the regular solution is given after setting Eq.~\eqref{dS+dm+} to zero:
\be{dT/dm:reg}
\C^{\rm reg} = \left(\frac{\partial M_+}{\partial T_+}\right)_{r_+}
= \frac{2 r_+}{T_+}\frac{v_+ }{f_+}. 
\ee
The value of $\C$ for self-gravitating regular sphere of radiations is positive definite in the region with $u_+ \to -\infty$ and changes signature when a solution curve crosses the lines $P$ and $H$.
$\C$ diverges and goes to zero when the solution curve intersects $P $ and $H$, respectively.

To obtain $\C$ for a general case with $r_- \neq 0$, the effect of $S_-$ should also be taken into account. 
Equating Eq.~\eqref{dS/dm+} by using Eqs.~\eqref{dv-:C}, \eqref{C:local}, \eqref{S:rad}, \eqref{dS:duv}, \eqref{dS+dm+} and using  $\left(\frac{\partial S_-}{\partial M_+}\right)_{r_\pm,M_-} 
=\left(\frac{\partial v_-}{\partial M_+}\right)_{r_\pm} 
	\left(\frac{\partial S_-}{\partial v_-}\right)_{r_-,u_-} $, 
we get
\bea  \label{Clocal:gen}
\C  =\frac{2r_+ v_+}{T_+f_+} \frac{1- \A}{1 +f_+^{-2}\, \A}, \qquad 
	\mathfrak{A} \equiv \sqrt{\frac{r_-}{r_+}} \frac{A_+}{A_-}, 
\eea
where we use Eqs.~\eqref{beta+}, \eqref{beta}, \eqref{beta-}, and 
\bea \label{Apm}
A_\pm \equiv  A(u_\pm,v_\pm), \qquad 
A(u,v) \equiv
   	\frac{v^{3/4} }{\chi }	
		\frac{f}{(2v -u)^2(1+f^2)} 
	= \frac{v^{3/4} \chi }{2v-u} \frac{F}{F^2 + G^2}.
\eea   
Note that the function $(2v-u) A(u,v)$ is a regular function on the whole range of physical interest other than the point $\mathcal{R}$, where $\mathcal{R}$ corresponds to the asymptotic infinity $r\to \infty$ of all solution curves. 
It vanishes on the lines $P$ and $v=0$.
The function $A(u,v)$ is positive definite in the region with $u \to -\infty$ and changes signature when a solution curve crosses the lines $P$ and $H$.
In the limit $r_- \to r_+$, the value of $\C$ goes to zero as expected.
When $\A \to 0$, the value of $\C$ is formally the same as that of the regular sphere in Eq.~\eqref{dT/dm:reg}.

Given $A$, the function $B$ can be determined by using the explicit value of $\C$ in Eq.~\eqref{Clocal:gen}. 
Equation~\eqref{du=0} gives
\be{B}
\sqrt{\frac{r_-}{r_+}} \frac{ f_-B_-}{f_+B_+} = \frac{ (2v_--u_-)A_-}{(2v_+ - u_+)A_+},
\ee
where $A_\pm \equiv A(u_\pm, v_\pm)$ is given in Eq.~\eqref{Apm}.
$B(u,v)$ must be a local function of $(u,v)$. 
Therefore, Eq.~\eqref{B} determines $B(u,v)$ up to a proportionality constant which is a function of $\nu$ only, 
\be{B:fA}
 B(u,v) = \alpha_\nu \sqrt{\frac{r_H}{r}} \frac{(2v-u) A(u,v)}{ f(u,v)}
 	=\alpha_\nu \sqrt{\frac{r_H}{r}} \frac{v^{3/4} \chi G}{
		F^2+ G^2} .
\ee
Here $\nu$ and $r/r_H = e^\xi$ are implicitly dependent on $u$ and $v$. 
It goes to zero on $H$ as expected in Eq.~\eqref{fBH}.
$B(u,v)$ diverges on $\mathcal{R}$.
The proportionality constant $\alpha_\nu$ can be fixed by using Eq.~\eqref{fBH}, after choosing $(u_+,v_+) = (u_H,v_H)$ and $(u_-,v_-) = (u,v)$, to be
\be{alpha nu}
\alpha_\nu =\lim_{r\to r_H}\frac{1}{ (2v-u) A} 
 	= \frac{2^{3/4}} {3}
		\frac{(7\nu-4)(1-\nu)^{1/4}}{\nu^{1/2}}.
\ee
Note that $\alpha_\nu $ is negative definite because $\nu$ is restricted to be $0 < \nu \leq \nu_r < 4/7$.


\begin{thebibliography}{99}

\bibitem{Landau1980}
L. Landau and E.M. Lifshitz, {\it Statiscal Physics}, 3rd ed. (Pergamon Press, New York), 21 (1980).

\bibitem{Lynden-Bell}
D. Lynden-Bell and R. Wood, 
Mon. Not. R. Astr. Soc. {\bf 138}, 495 (1968).

\bibitem{Sorkin:1981wd} 
  R.~D.~Sorkin, R.~M.~Wald and Z.~J.~Zhang,
  Gen.\ Rel.\ Grav.\  {\bf 13}, 1127 (1981).

\bibitem{Schmidt:1999tr} 
  H.~J.~Schmidt and F.~Homann,
  Gen.\ Rel.\ Grav.\  {\bf 32}, 919 (2000)
  [gr-qc/9903044].
  
\bibitem{Pavon1988}
D.~Pavon and P. T.~Landsberg, 
Gen.\ Rel.\ Grav.\ {\bf 20}, 457 (1988).

\bibitem{Chavanis:2007kn} 
  P.~H.~Chavanis,
  Astron.\ Astrophys.\  {\bf 483}, 673 (2008)
  [arXiv:0707.2292 [astro-ph]].
  
\bibitem{Chavanis:2001hd} 
  P.~H.~Chavanis,
  Astron.\ Astrophys.\  {\bf 381}, 340 (2002)
  doi:10.1051/0004-6361:20011438
  [astro-ph/0103159].

\bibitem{Chavanis:2001ib} 
  P.~H.~Chavanis, C.~Rosier and C.~Sire,
  Phys.\ Rev.\ E {\bf 66}, 036105 (2002)
  doi:10.1103/PhysRevE.66.036105
  [cond-mat/0107345].
  


\bibitem{Schiffer:1989et} 
  M.~Schiffer and J.~D.~Bekenstein,
  Phys.\ Rev.\ D {\bf 39}, 1109 (1989).
  doi:10.1103/PhysRevD.39.1109
  
\bibitem{Hod:1999as} 
  S.~Hod,
  gr-qc/9901035.
  
\bibitem{Sorkin:1997ja} 
  R.~D.~Sorkin,
  gr-qc/9705006.
 
\bibitem{Gao:2016trd} 
  S.~Gao,
  Springer Proc.\ Phys.  {\bf 170}, 359 (2016).
  doi:10.1007/978-3-319-20046-043
  
\bibitem{Gao:2011hh} 
  S.~Gao,
  Phys.\ Rev.\ D {\bf 84}, 104023 (2011)
  [Phys.\ Rev.\ D {\bf 85}, 027503 (2012)]
  doi:10.1103/PhysRevD.84.104023, 10.1103/PhysRevD.85.027503
  [arXiv:1109.2804 [gr-qc]].

\bibitem{Fang:2013oka} 
  X.~Fang and S.~Gao,
  Phys.\ Rev.\ D {\bf 90}, no. 4, 044013 (2014)
  doi:10.1103/PhysRevD.90.044013
  [arXiv:1311.6899 [gr-qc]].
  
\bibitem{Lemos:2007ys} 
  J.~P.~S.~Lemos,
  arXiv:0712.3945 [gr-qc].


\bibitem{Anastopoulos:2013xdk} 
  C.~Anastopoulos and N.~Savvidou,
  Class.\ Quant.\ Grav.\  {\bf 31}, 055003 (2014)
  doi:10.1088/0264-9381/31/5/055003
  [arXiv:1302.4407 [gr-qc]].

\bibitem{Bousso:2002ju} 
  R.~Bousso,
  Rev.\ Mod.\ Phys.\  {\bf 74}, 825 (2002)
  doi:10.1103/RevModPhys.74.825
  [hep-th/0203101].


\bibitem{Page:2013mqa} 
  D.~N.~Page,
  JCAP {\bf 1406}, 051 (2014)
  doi:10.1088/1475-7516/2014/06/051
  [arXiv:1306.0562 [hep-th]].
  
\bibitem{Page:1985em} 
  D.~N.~Page and K.~C.~Phillips,
  Gen.\ Rel.\ Grav.\  {\bf 17}, 1029 (1985).
  doi:10.1007/BF00774206
  
\bibitem{Vaganov:2007at} 
  V.~Vaganov,
  arXiv:0707.0864 [gr-qc].
  
\bibitem{Gentle:2011kv} 
  S.~A.~Gentle, M.~Rangamani and B.~Withers,
  JHEP {\bf 1205}, 106 (2012)
  doi:10.1007/JHEP05(2012)106
  [arXiv:1112.3979 [hep-th]].

\bibitem{Pesci:2006sb} 
  A.~Pesci,
  Class.\ Quant.\ Grav.\  {\bf 24}, 2283 (2007)
  doi:10.1088/0264-9381/24/9/009
  [gr-qc/0611103].
    
\bibitem{Anastopoulos:2011av} 
  C.~Anastopoulos and N.~Savvidou,
  Class.\ Quant.\ Grav.\  {\bf 29}, 025004 (2012)
  [arXiv:1103.3898 [gr-qc]].

\bibitem{Zurek}
W. H. Zurek and D. N. Page, Phys.\ Rev.\ D {\bf 29}, 628 (1984).

\bibitem{Anastopoulos:2014zqa} 
  C.~Anastopoulos and N.~Savvidou,
  JHEP {\bf 1601}, 144 (2016)
  doi:10.1007/JHEP01(2016)144
  [arXiv:1410.0788 [gr-qc]].

\bibitem{Kim:2016jfh} 
  H.~C.~Kim,
  arXiv:1601.02720 [gr-qc].


\end{thebibliography}
\end{document}